\documentclass[nofootinbib]{revtex4}
\usepackage{latexsym}
\usepackage{amssymb}
\usepackage{amsfonts}
\usepackage{amsmath}
\usepackage[dvips]{graphicx}
\usepackage{bm}

\def\ber{\begin{eqnarray}}
\def\eer{\end{eqnarray}}
\def\beq{\begin{equation}}
\def\eeq{\end{equation}}

\begin{document}

\title{Pulsars as celestial beacons to detect the motion of the Earth}

\begin{abstract}
In order to show the principle viability of a recently proposed relativistic positioning method based on the use of pulsed signals from sources at infinity, we present an application example reconstructing the world-line of an idealized Earth in the reference frame of distant pulsars. The method considers the null four-vectors built from the period of the pulses and the direction cosines of the propagation from each source. Starting from a simplified problem (a receiver at rest) we have been able to calibrate our procedure, evidencing the influence of the uncertainty on the arrival times of the pulses as measured by the receiver, and of the numerical treatment of the data. The most relevant parameter turns out to be the accuracy of the clock used by the receiver. Actually the uncertainty used in the simulations combines both the accuracy of the clock and the fluctuations in the sources. As an evocative example the method has then been applied to the case of an ideal observer moving as a point on the surface of the Earth. The input have been the simulated arrival times of the signals from four pulsars at the location of the Parkes radiotelescope in Australia. Some substantial simplifications have been made both excluding the problems of visibility due to the actual size of the planet, and the behaviour of the sources. A rough application of the method to a three days run gives a correct result with a poor accuracy. The accuracy is then enhanced to the order of a few hundred meters if a continuous set of data is assumed. The method could actually be used for navigation across the solar system and be based on artificial sources, rather than pulsars. The viability of the method, whose additional value is in the self-sufficiency, i.e. independence from any control from other operators, has been confirmed.
\end{abstract}

\author{Matteo Luca Ruggiero}
\email{matteo.ruggiero@polito.it}
\affiliation{Dipartimento di Fisica, Politecnico di Torino, Corso Duca degli Abruzzi 24, 10129 Torino, Italy}
\affiliation{INFN, Sezione di Torino, Via Pietro Giuria 1, 10125 Torino, Italy}

\author{Emiliano Capolongo}
\email{emiliano.capolongo@polito.it}
\affiliation{Dipartimento di Fisica, Politecnico di Torino, Corso Duca degli Abruzzi 24, 10129 Torino, Italy}
\affiliation{INFN, Sezione di Torino, Via Pietro Giuria 1, 10125 Torino, Italy}

\author{Angelo Tartaglia}
\email{angelo.tartaglia@polito.it}
\affiliation{Dipartimento di Fisica, Politecnico di Torino, Corso Duca degli Abruzzi 24, 10129 Torino, Italy}
\affiliation{INFN, Sezione di Torino, Via Pietro Giuria 1, 10125 Torino, Italy}

\date{\today }


\maketitle

\section{Introduction}\label{sec:intro}

Soon after the discovery of pulsars, it was suggested that they could have been used as stellar beacons for spacecraft navigation in the Solar System and beyond \cite{oldiepuls}. Today there are proposals focusing on the use of X-ray pulsars for navigation  \cite{xrays1,xrays2}; they are based on the accurate measurement of the times of arrival of pulses or phase differences, in order to determine the position of the spacecraft.
In  previous papers \cite{corea,pulsararc10} we operationally described how a relativistic positioning system can be build using electromagnetic signals emitted by pulsating sources, such as pulsars, thanks to the use of emission coordinates \cite{coll3}. The simplest way of understanding how emission coordinates work, is to
consider four emitting clocks, in motion through space while broadcasting their
proper times. The intersections of the past lightcone of an event with the
world-lines of the emitting clocks can be labeled with the proper times of
emission along the world-lines of the emitters: these proper times are the
emission coordinates of the given event. We showed that, by receiving pulses from a set of different
sources whose positions in the sky and periods are assumed to be known,
it is possible to determine  the user's coordinates and spacetime trajectory,
in the reference frame where the sources are at rest. In doing so, the
phases of the received pulses play the role of emission coordinates.  In particular, we developed a  procedure that can be used to determine the user's trajectory by assuming that its world-line is a straight line during a proper time interval corresponding to the reception of a limited number of pulses, which means that the effects of the
acceleration are negligibly small. Our approach is based on the use of null frames in flat Minkowski spacetime, but we discussed its possible application to actual physical events provided that suitable approximations hold true.

In this paper we want to practically develop an application of our method. In order to do so we actually imagine that our sources are four millisecond pulsars, then, simulating the arrival times of their signals, we show how the worldline of the receiver is reconstructed. First, for a sort of calibration, we imagine to have an observer at rest with respect to "fixed" stars and reconstruct his, in this case, trivial spacetime trajectory, with the uncertainties implied by the procedure. Then,  making use of the TEMPO2 software \cite{tempo2}, a pulsar-timing package that simulates the times of arrival of pulses at a given location on the Earth, we determine the trajectory of that location in spacetime, due to the combined motion of the Earth around the Sun and to its daily rotation. Eventually, we compare the reconstructed world-line with the one obtained by the ephemerides, and discuss the achieved accuracy. Both examples are meant as feasibility tests of the whole process; the problems with a real positioning system are discussed in the conclusion.

The paper is organized as follows: in Section \ref{sec:method} we outline the theoretical framework, then the results of our simulations are given in Section \ref{sec:numsim}. Conclusions are given in Section \ref{sec:conc}.

\section{The Method} \label{sec:method}
In this Section, we give an outline of the method that we will apply  to reconstruct the spacetime trajectory.

To begin with, let us introduce the basic null frame which  allows to determine a worldline. In order to define a four-dimensional spacetime frame based on the use of null-coordinates,  at least four sources of electromagnetic signals are needed. To fix ideas, we suppose to pick up a set of four pulsars,  provided their periods and angular positions in the sky are known: namely,  each of these sources is characterized by the frequency of its periodic signals and by their propagation directions in space; the emitters  are supposed to be at rest, in a given reference frame,  at spatial infinity (hence,  their signals can be thought of as plane waves)

In the reference frame where the sources are at rest we
associate to each of them a null four-vector\footnote{%
Arrowed bold face letters like $\vec{\mathbf{x}}$ refer to spatial vectors,
while boldface letters like $\bm f$ refer to four-vectors; Greek indices
refer to spacetime components, while Latin letters label the sources.} $\bm f
$ whose Cartesian contravariant components are given by
\begin{equation}
f^{\mu } \doteq \frac{1}{c T}(1,\vec{\mathbf{n}}),  \label{eq:deff}
\end{equation}
$T$ is the (proper) signal period, and $\vec{%
\mathbf{n}}$ is the unit vector describing the direction of propagation in
the given frame.
A receiver counts the periodic electromagnetic signals coming from the sources and measures the proper time intervals between successive arrivals.
To each spacetime event defined by the position four-vector
\begin{equation}
\bm{r}\doteq (ct,\vec{\mathbf{x}}),  \label{eq:defr}
\end{equation}%
it is possible to associate the scalar function $X(\bm r)$
\begin{equation}
X(\bm r)\doteq \bm f\cdot \bm r,  \label{eq:defX}
\end{equation}%
where the dot stays for Minkowski scalar product. The scalar $X$ might be
thought of as the phase difference of the wave described by $\bm f$ with
respect to its value at the origin of the coordinates, where $\bm r= \bm 0$.  Given  four emitters,  the
four wave four-vectors $\{\bm f_{(a)},\bm f_{(b)},\bm f_{(c)},\bm f_{(d)}\}$
in the form (\ref{eq:deff}) constitute the null frame or null tetrad (where $a,b,c,d$ label the sources). In particular, at any event $\bm r$ the phases
\begin{equation}
X_{(N)}\doteq \bm f_{(N)}\cdot \bm r,\quad N=a,b,c,d\   \label{eq:defXN}
\end{equation}%
are defined. Then, starting from the definition of the symmetric matrix
\begin{equation}
\eta _{(M)(N)}=\bm f_{(M)}\cdot \bm f_{(N)},  \label{eq:defgab}
\end{equation}
and  its inverse $\eta ^{(P)(N)}$ (such that $\eta _{(M)(P)}\eta ^{(P)(N)}=\delta _{(M)}^{(N)}$), we define the vectors
\begin{equation}
\bm f^{(N)}=\eta ^{(N)(M)}\bm f_{(M)},  \label{eq:fcontraN}
\end{equation}
Eventually, as we showed in \cite{pulsararc10}, the position four-vector is expressed in the form
\begin{equation}
\bm r = X_{(a)}\bm f^{(a)}+X_{(b)}\bm f^{(b)}+X_{(c)}\bm f^{(c)}+X_{(d)}\bm %
f^{(d)}.  \label{eq:defXzero3}
\end{equation}
where the phases $X_{(N)}$ play the role of coordinates with respect to the null frame; in other words, eq. (\ref{eq:defXzero3}) shows that it is possible to obtain the
coordinates of event $\bm r$, in terms of the measured phases and to reconstruct the  world-line of the receiver. However, in actual situations, the received signals consist in a series of pulses so that, in general,  the values of $X_{(N)}(\bm r)$ are not directly observable.  In what follows we address this issue and describe how phases can be empirically determined, provided that some assumptions hold true. \\

To begin with, we  call \textquotedblleft reception\textquotedblright\ the
event corresponding to the arrival of a pulse from one of the sources. As a
consequence, an arbitrary reception event can be written in the form  (\ref{eq:defXzero3})
with
\begin{eqnarray}
X_{(a)} &=&n_{(a)}+p,  \label{eq:defp} \\
X_{(b)} &=&n_{(b)}+q,  \label{eq:defq} \\
X_{(c)} &=&n_{(c)}+s,  \label{eq:defs} \\
X_{(d)} &=&n_{(d)}+w,  \label{eq:defw}
\end{eqnarray}%
where we have expressed the phases $X_{(N)}$ in terms of an integer $n_{(N)}$%
, describing the sequence of cycles of signals, and a fractional value: e.g.
$p$ means a fractional value of the cycle in $X^{(a)}$, and the equivalent
holds for $q,s,w$, where $0<p,q,s,w$. This amounts to saying that, at any reception event, in eqs. (\ref{eq:defp})-(\ref{eq:defw}%
), only one of the $p,q,s,w$ is in general null. Once we choose an
arbitrary origin, we may count the pulses in order to measure the $n_{(N)}$,
but we have no direct means to measure the fractional values $p,q,s,w$.
However, simple geometric considerations allow to introduce a procedure to
determine these values: first,  we suppose that the acceleration of the receiver is
small during a limited series of reception events, so that we may identify
the user's  world-line with a straight line; then, we  suppose that,
by means of its own clock, the receiver can measure the proper time interval $%
\tau _{ij}$ between the i-th and j-th arrivals. With these assumptions it is possible to determine the fractional values $p,q,s,w$. To this end, let us
consider two sequences\footnote{%
They may be subsequent or not, provided the total time span does not spoil
the hypothesis of linearity of the  world-line.} of arrival times from the
sources: they consist of eight reception events, each of them in the form
\begin{equation}
\bm r_{j}=X_{(a)j}\bm f^{(a)}+X_{(b)j}\bm f^{(b)}+X_{(c)j}\bm %
f^{(c)}+X_{(d)j}\bm f^{(d)},    j=1,..,8,  \label{eq:arrivalj}
\end{equation}%
where $X_{(N)j}$'s are expressions like (\ref{eq:defp}--\ref{eq:defw}).  We arrange the events as follows: $\bm r_{1}$ is the arrival of the generic signal from pulsar (or any other equivalent source) \textquotedblleft a\textquotedblright , $\bm r_{2}$
is the arrival of the first signal of pulsar \textquotedblleft
b\textquotedblright\ after $\bm r_{1}$, $\bm r_{3}$ is the  arrival of
the first signal of pulsar \textquotedblleft c\textquotedblright\ after $\bm %
r_{1}$, and $\bm r_{4}$ is the  arrival of the first signal of pulsar
\textquotedblleft d\textquotedblright\ after $\bm r_{1}$ (the sources are
ordered from largest (\textquotedblleft a\textquotedblright ) to shortest
(\textquotedblleft d\textquotedblright ) period); $\bm r_{5}$ is the arrival
of the first \textquotedblleft a\textquotedblright  signal of the second sequence, and
so on.  The flatness hypothesis allows to write the displacement four-vector
between two reception events in the form
\begin{equation}
\bm r_{ij}\doteq \bm r_{i}-\bm r_{j}=\left( X_{(N)i}-X_{(N)j}\right) \bm %
f^{(N)}\doteq \Delta X_{(N)ij}\bm f^{(N)}.  \label{eq:deltarij}
\end{equation}%
In fact, the assumption that the  world-line of the receiver is straight during
a limited number of paces of the signals can be used also to provide
further information. Let us consider three successive reception
events i,j,k; we have
\begin{equation}
\bm r_{ji}=\Delta X_{(N)ji}\bm f^{(N)},\quad \quad \bm r_{kj}=\Delta
X_{(N)kj}\bm f^{(N)}.  \label{eq:deltarjirkj}
\end{equation}%
The straight-line hypothesis allows us to write
\begin{equation}
\frac{\tau _{ji}}{\tau _{kj}}=\frac{\Delta X_{(a)ji}}{\Delta X_{(a)kj}}=%
\frac{\Delta X_{(b)ji}}{\Delta X_{(b)kj}}=\frac{\Delta X_{(c)ji}}{\Delta
X_{(c)kj}}=\frac{\Delta X_{(d)ji}}{\Delta X_{(d)kj}},  \label{eq:rapporti1}
\end{equation}%
where $\tau _{ji}$, $\tau _{kj}$ are the proper times elapsed between the
i-th and j-th, and j-th and k-th reception events, respectively. These
relations enable us to obtain the values we are interested in: in fact, we
may write eq. (\ref{eq:defXzero3}) in the form
\begin{equation}
\bm r_{j}=X_{(a)j}\bm f^{(a)}+X_{(b)j}\bm f^{(b)}+X_{(c)j}\bm %
f^{(c)}+X_{(d)j}\bm f^{(d)},j=1,..,8,  \label{eq:arrivalj1}
\end{equation}%
where $X_{(N)j}$ is given by
\begin{widetext}
\beq
X_{(N)i}=\left(\begin{array}{c|c|c|c} n_{1}^{(a)} & n^{(b)}_{1}+q_{1} & n^{(c)}_{1}+s_{1} & n^{(d)}_{1}+w_{1} \\\hline n^{(a)}_{2}+p_{2} & n^{(b)}_{2} & n^{(c)}_{2}+s_{2} & n^{(d)}_{2}+w_{2} \\\hline n^{(a)}_{3}+p_{3} & n^{(b)}_{3}+q_{3} & n^{(c)}_{3} & n^{(d)}_{3}+w_{3} \\\hline n^{(a)}_{4}+p_{4} & n^{(b)}_{4}+q_{4} & n^{(c)}_{4}+s_{4} & n^{(d)}_{4} \\\hline n^{(a)}_{5} & n^{(b)}_{5}+q_{5} & n^{(c)}_{5}+s_{5} & n^{(d)}_{5}+w_{5} \\\hline n^{(a)}_{6}+p_{6} & n^{(b)}_{6} & n^{(c)}_{6}+s_{6} & n^{(d)}_{6}+w_{6} \\\hline n^{(a)}_{7}+p_{7} & n^{(b)}_{7}+q_{7} & n^{(c)}_{7} & n^{(d)}_{7}+w_{7} \\\hline n^{(a)}_{8}+p_{8} & n^{(b)}_{8}+q_{8} & n^{(c)}_{8}+s_{8} & n^{(d)}_{8}\end{array}\right) \label{eq:tableXNi}
\eeq
and the the fractional values are expressed in terms of observed quantities by:
\beq
p_{1}=0, \quad q_{1}=n^{(b)}_{2}-n^{(b)}_{1}-\left(n^{(b)}_{6}-n^{(b)}_{2} \right) \frac{\tau_{21}}{\tau_{62}}, \quad s_{1}=n^{(c)}_{3}-n^{(c)}_{1}-\left(n^{(c)}_{7}-n^{(c)}_{3} \right) \frac{\tau_{31}}{\tau_{73}},\quad w_{1}=n^{(d)}_{4}-n^{(d)}_{1}-\left(n^{(d)}_{8}-n^{(d)}_{4} \right) \frac{\tau_{41}}{\tau_{84}}, \label{eq:ext11}
\eeq
\beq
p_{2}=n^{(a)}_{1}-n^{(a)}_{2}+\left(n^{(a)}_{5}-n^{(a)}_{1} \right) \frac{\tau_{21}}{\tau_{51}}, \quad q_{2}=0, \quad s_{2}=n^{(c)}_{3}-n^{(c)}_{2}+\left(n^{(c)}_{7}-n^{(c)}_{3} \right) \frac{\tau_{23}}{\tau_{73}},\quad w_{2}=n^{(d)}_{4}-n^{(d)}_{2}+\left(n^{(d)}_{8}-n^{(d)}_{4} \right) \frac{\tau_{24}}{\tau_{84}}, \label{eq:ext12}
\eeq
\beq
p_{3}=n^{(a)}_{1}-n^{(a)}_{3}+\left(n^{(a)}_{5}-n^{(a)}_{1} \right) \frac{\tau_{31}}{\tau_{51}}, \quad q_{3}=n^{(b)}_{2}-n^{(b)}_{3}-\left(n^{(b)}_{6}-n^{(b)}_{2} \right) \frac{\tau_{23}}{\tau_{62}}, \quad s_{3}=0,\quad w_{3}=n^{(d)}_{4}-n^{(d)}_{3}+\left(n^{(d)}_{8}-n^{(d)}_{4} \right) \frac{\tau_{34}}{\tau_{84}}, \label{eq:ext13}
\eeq

\beq
p_{4}=n^{(a)}_{1}-n^{(a)}_{4}+\left(n^{(a)}_{5}-n^{(a)}_{1} \right) \frac{\tau_{41}}{\tau_{51}}, \quad q_{4}=n^{(b)}_{2}-n^{(b)}_{4}-\left(n^{(b)}_{6}-n^{(b)}_{2} \right) \frac{\tau_{24}}{\tau_{62}}, \quad s_{4}=n^{(c)}_{3}-n^{(c)}_{4}-\left(n^{(c)}_{7}-n^{(c)}_{3} \right) \frac{\tau_{34}}{\tau_{73}},\quad w_{4}=0, \label{eq:ext14}
\eeq
\end{widetext}and so on.

The procedure that we have just described allows to give an operational meaning to the phases (\ref{eq:defXN}) and, hence, to the determination of the position four-vector, through (\ref{eq:defXzero3}). By simply measuring  proper times and  considering sequences of octets, it is possible to reconstruct the whole receiver's world-line.

In what follows, we test the reliability of this procedure; in particular, we  consider different situations where four emitters send pulsed signals to the receiver. We are going to show how, by measuring the proper time intervals between reception events, it is possible to reconstruct the receiver's trajectory in spacetime, once the emission directions and frequencies of the pulsating signals are known.

\section{Numerical Simulations} \label{sec:numsim}

In order to test the procedure described above, it is necessary to define the null frame, that is to say the basis four-vectors in the form (\ref{eq:deff}) for each source. In other words, we need to know the positions of the sources and their periods. Then, in order to apply the procedure, we need  the arrival times of the pulses, as measured by the receiver. Our purpose is to demonstrate how the system works in practice, but we have no actual device at hands so we may follow two strategies: a) simulate the sources, giving them an arbitrary position in the sky and an arbitrary periodicity, then somehow mimicking the uncertainties associated with real sources; b) choose, as an example, four real millisecond pulsars with the data we find in the literature. In practice the difference between the two approaches is not really important, since the second choice is only nominally different from the first, so we decided to use the parameters of four real pulsars as they are listed in Table \ref{tab:table1}.

Next we proceed in two steps:
\begin{itemize}
\item a) we numerically simulate the reception events of the pulses, starting from the knowledge of the receiver's trajectory and the definition of the null frame; this is in particular applied to the case of an observer at rest with "fixed" stars.
\item b) we use a software which simulates the arrival times of the pulses received at a given terrestrial location, emitted by our set of pulsars; by this way we try and reconstruct the worldline of the receiver at the chosen position on Earth.
  \end{itemize}

In both cases a previous knowledge is used to generate a series of arrival times as they would be sensed by the receiver's antenna. In the first case, the output of our algorithm is compared with the receiver's worldline, that we know is a straight line, and we can test how any uncertainty present in the input arrival times is transferred to the outcome. In the second case, the simulator generates sequences of arrival times as they would be obtained at an antenna and we use them, applying our algorithm in order to rebuild the motion of the Earth or, more correctly, the trajectory of a terrestrial location where the pulses would be received, which moves because of the daily rotation and the motion of the Earth around the Sun: this trajectory is then compared to the one obtained by the ephemerides. More precisely we reconstruct the motion of the receiver with respect to the "fixed" stars, assuming as the origin the event where the reception has started: in a sense we produce a self-positioning. The position of the initial event with respect to any given reference frame must be known by other means.

\begin{table}[here]
\begin{center}
\medskip
\begin{tabular}{ccccc}
& & & & \\ \hline
Pulsar      &  T (ms) & Elong $(^{\circ})$ & Elat $(^{\circ})$ \\ \hline
J1730-2304  & $8.123$ & $263.19$      & $0.19$  \\
J2322+2057	& $4.808$ & $0.14$        & $22.88$ \\
B0021-72N	& $3.054$ & $311.27$      & $-62.35$  \\
B1937+21	& $1.558$ & $301.97$      & $42.30$   \\  \hline
\end{tabular}
\end{center}
\caption{The parameters of the four pulsars we chose are listed as they were taken from the ATNF Pulsar Catalogue. The basis four-vectors are obtained after computing the direction cosines from the ecliptic coordinates; then use is made of the formula $\bm f _{(N)} =\frac{1}{c T_{(N)}}(1,\vec{\mathbf{n}}_{(N)})$, for $N=a,\ b,\ c,\ d$. Both the periods and the direction cosines are assumed to be known with an accuracy limited by the numerical precision only.}
\label{tab:table1}
\end{table}

\subsection{User at rest} \label{ssec:rest}

As written above, the first step is to consider the simple case of a receiver at rest in the reference frame of the sources: this can be thought of as a sort of calibration of our procedure.
We use the parameters of the four real pulsars listed in Table \ref{tab:table1}, chosen from the ATNF Pulsar Catalogue \cite{catalogo}. As for the positions and periods, in the simulation they have an accuracy limited by the numerical precision only; actually the real uncertainty would produce here a systematic error which is irrelevant for our purposes. In any case, we introduced a Gaussian error with a 1 ns $\sigma$ in the determination of the times of arrival in order to account both for the accuracy of the receiver's clock and for the uncertainty in the behaviour of the sources, due to the fluctuation of their periods. If we want to be realistic we should recognize that the chosen figure is rather optimistic or at least it would require a rather long integration time. Given the example we are working out here (observer at rest) the length of the integration time is indeed no problem, however we must also recall that our aim is simply to see how the assumed uncertainty is reflected in the final result, without worrying about the spectral composition of the noise or other details. For the same reasons we did not take into account either the proper motions of the sources or the natural decay of their periods, considering that these factors are irrelevant in such short times as the ones relevant for our simulation; if needed they could easily be included.

The arrival times of the pulses are obtained numerically, by considering the intersection of the receiver's world line
\begin{equation}
\rho \equiv \left\{\begin{array}{ccc}x(\lambda) &=& 0 \\y(\lambda) &=& 0 \\z(\lambda) &=& 0 \\t(\lambda) & = &   \lambda\end{array}\right. \label{eq:rholambda1}
\end{equation}
with the constant phase hyperplanes from the sources
\beq
\bm f_{(N)} \cdot \bm r = \mathcal{N}_{(N)}, \quad N=a,\ b,\ c,\ d \label{eq:elicapiani1}
\eeq
whose zeros have been arbitrarily chosen.

The arrival times have been ordered according to the procedure described in Section \ref{sec:method} and, then, they have been  used to compute the phases at the reception events. Eventually, the reconstructed receiver trajectory $\bar \rho $ has been obtained in terms of these phases, according to eq. (\ref{eq:defXzero3}):
\begin{equation}
\bm r [ {i}] = X[ {i}]_{(a)}\bm f^{(a)}+X[ {i}]_{(b)}\bm f^{(b)}+X[ {i}]_{(c)}\bm f^{(c)}+X[ {i}]_{(d)}\bm %
f^{(d)}.   \label{eq:defXzero3elica}
\end{equation}
where $[  i]$ is an index labeling the i-th reception event.  In particular, from (\ref{eq:defXzero3elica}) we obtain the Cartesian components $\bar t, \bar x, \bar y, \bar z$ for each reception event.

\begin{figure}[here]
\begin{center}
\includegraphics[scale=.70]{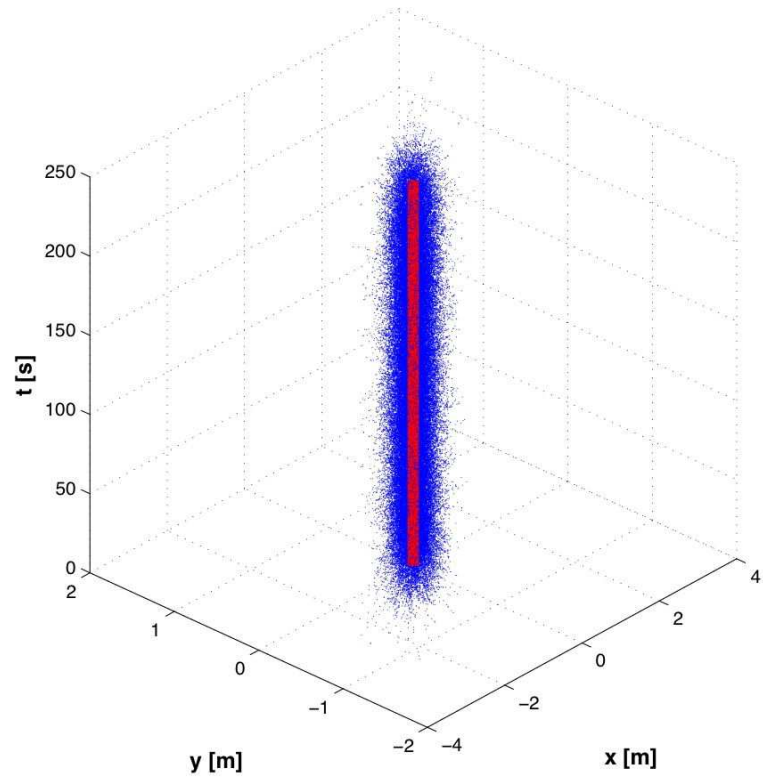}
\end{center}
\caption{Reconstructed world-line of an user at rest. The straight vertical line is the expected result; the scattered blue points are the reconstructed positions obtained applying our method with a nanosecond Gaussian noise in the input. The RMS dispersion is of the order of less than 40 cm. The greater dispersion of the $x$ coordinates with respect to $y$ is due to the anisotropic distribution of the sources in the sky.}
\label{fig:resta}
\end{figure}

\begin{figure}[here]
\begin{center}
\includegraphics[scale=.70]{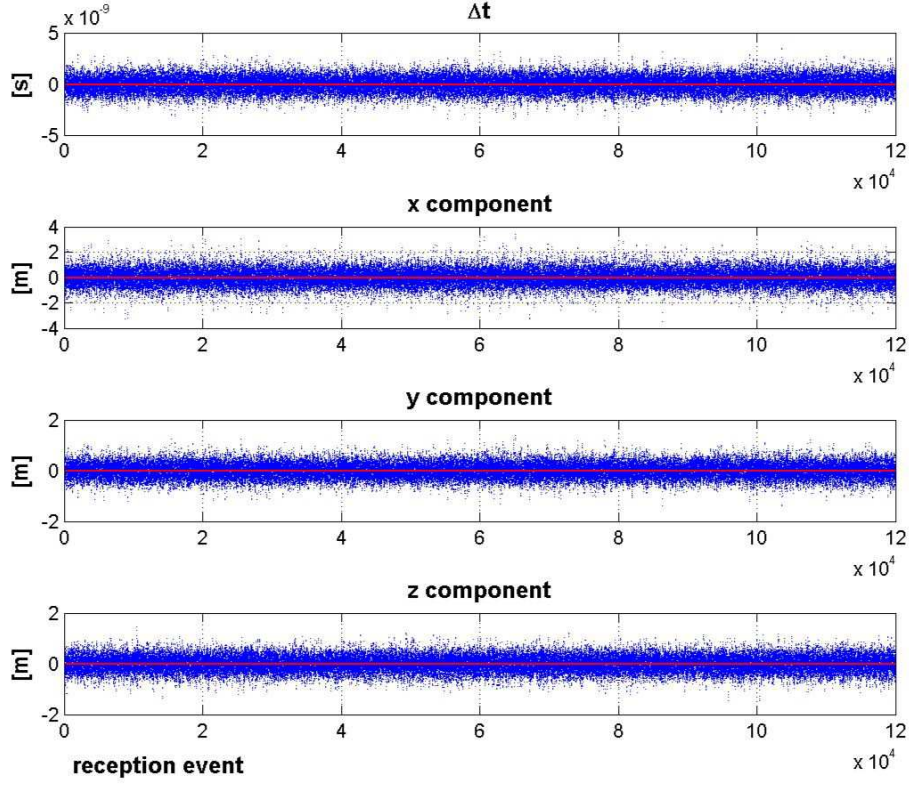}
\end{center}
\caption{The same as for Figure \ref{fig:resta}, but for each coordinate separately; $\Delta t \doteq \bar t-t$ is the difference between the reconstructed time component and the expected one. As it can be seen the accuracy is of the same order as the one on the arrival times. The greater dispersion of the $x$ coordinates with respect to the other two is due to the anisotropic distribution of the sources in the sky.}
\label{fig:restb}
\end{figure}

The comparison between the analytic spacetime trajectory $ \rho$ and the reconstructed one $\bar \rho$ is given in Figure \ref{fig:resta}, in a bidimensional-plus-time view; as it can be seen the dispersion of the reconstructed positions is contained within one meter or less.  In Figure \ref{fig:restb} the components of the reconstructed trajectory are given, as functions of the sequence of the reception events. It is possible to see that the $x$-component has a bigger dispersion than the other two; this is due to the fact that the sources are not isotropically located in the sky (the ideal configuration should be tetrahedron-like, which cannot be in the real world).
 The RMS deviation of the reconstructed positions with respect to the ideal ones (vertical bar in the origin) is less than 0.40  m; this is consistent with the 1 ns uncertainty that we assumed for the arrival times.

\

\subsection{Reconstructing the motion of the Earth} \label{ssec:terra}

The next step, as announced, is to use our method for a less trivial and a bit more realistic situation. It would be interesting to use real data, i.e. a sequence of arrival times from known pulsars, as measured at a radiotelescope. On doing so, it would be possible to rebuild the motion of the radiotelescope, determined by the combined effect of the terrestrial rotation and the motion of the Earth around the Sun.

Actually, strictly speaking, our approach to the use of pulsating sources for positioning is defined in Minkowski spacetime and the sources are supposed to be ideal, i.e. at rest and emitting with a constant frequency. So, one could expect that it is not possible to use it to deal with a real situation where the gravitational field is present (such as it is the case of the Earth in the gravitational field of the Sun) and the sources are real pulsars. However, as discussed in \cite{pulsararc10},  the effects of the gravitational field,  the proper motion of the sources and the variability of their emission rate are small enough to be neglected for many actual physical situations, among which the one in which we are interested.

Hence, in principle, it  could be possible to define a null basis by taking four known millisecond pulsars (which have a highly regular emission rate, see e.g. \cite{kramer}), suitably distributed in the  sky (in an as much tetrahedral as possible configuration). The receiver could be one of the terrestrial radiotelescopes, now spread worldwide, which are able to point, hook up and follow the faint signals emitted by the chosen pulsars. If it were possible for a given radiotelescope to simultaneously receive four signals from four sources, then the application of our procedure could determine the spacetime trajectory of the radiotelescope. Unfortunately, for the moment, it is technically impossible for a given radiotelescope to simultaneously track four distinct pulsars, located at very different positions in the sky.\footnote{However, the new radio telescopes, such as the planned square kilometer array\cite{ska} (SKA), will be able to do this.}

Given the purpose of our exercise we may keep the exterior dressing of a real measurement, but using simulated arrival times.  To this aim we shall use TEMPO2 \cite{tempo2}, a specific software environment, widely used nowadays by the astronomers and astrophysicists studying pulsars, which enables to simulate the pulsar timing. In particular, the TEMPO2 plug-in ``\textit{fake}'' enables to simulate the time residuals expected from a given pulsar observation session. In fact, this code automatically  generates a set of times of arrival for a specific pulsar at a predefined location on the Earth surface (corresponding, for instance, to  a radiotelescope site), in a time window defined by the user, and starting from the transit time of the given pulsar through the local meridian (superior culmination point). It takes into account the contribution to timing of the gravitational field in the Solar System due to the Sun and the other bodies, and other kinematical effects (see e.g. \cite{straumann04}).  The possibility to add various types of error, in particular the Gaussian one or the red noise one (a timing noise that is actually negligible for most millisecond pulsars), to the times of arrival is also allowed.

Hence, we chose the four pulsars we already used for the previous exercise, introducing here a Gaussian 1 $\mu$s uncertainty. This is a conservative estimate of the error in the timing procedure, due both to the detection process and to the fluctuations of the sources. Then TEMPO2 has been for us the equivalent of an antenna where the sequences of pulses from our quartet of sources are received.

We of course know that TEMPO2 presupposes a complete knowledge of the position of the sources, which in turn is obtained assuming the motion of the observatory with respect to fixed stars is already given by other means. In practice we have a logic loop here, however in a real situation instead of TEMPO2 we would have a receiving antenna and the downstream processing of the data is insensitive to the real nature of the input. Furthermore, as already mentioned, ours is, strictly speaking, a self-positioning rather than an absolute positioning. Given three independent space directions in the sky (directions of the axes of the reference frame) the origin is assumed to coincide with the starting point of the positioning process. The location of that origin with respect to some global reference frame, such as the International
Celestial Reference System (ICRS), must be independently defined.

The arrival times have been simulated during a time interval of  about three days, at  a given position on the surface of the Earth, that is the one of the Parkes observatory in Australia.  In particular, we considered for each pulsar a set of about 28000 pulses, sampled out of the continuous sequence each 10 seconds. The duration of the simulation allows to evidence the actual motion of the observatory, due to the combined motion of the Earth around the Sun and of its daily rotation. The chosen pulsars define the null frame, and they are supposed to be at rest in the ICRS (where, in turn, the barycenter of the Solar System is at rest).

By applying the procedure described above, we have rebuilt the trajectory of the observatory. To make a comparison, we consider $\zeta$,  that is the trajectory of the observatory, as determined by the ICRS ephemerides having components $t,x,y,z$, while, as before, the reconstructed trajectory $\bar \zeta$ has been obtained  according to eq. (\ref{eq:defXzero3}):
\begin{equation}
\bm r [ i] = X[  i]_{(a)}\bm f^{(a)}+X[  i]_{(b)}\bm f^{(b)}+X[  i]_{(c)}\bm f^{(c)}+X[  i]_{(d)}\bm %
f^{(d)}.  \label{eq:defXzero3earth}
\end{equation}
where $[ i]$ is an index labeling the i-th reception event.  In particular, from (\ref{eq:defXzero3earth}) we obtain the Cartesian components $\bar t, \bar x, \bar y, \bar z$.

\begin{figure}[here]
\begin{center}
\includegraphics[scale=.70]{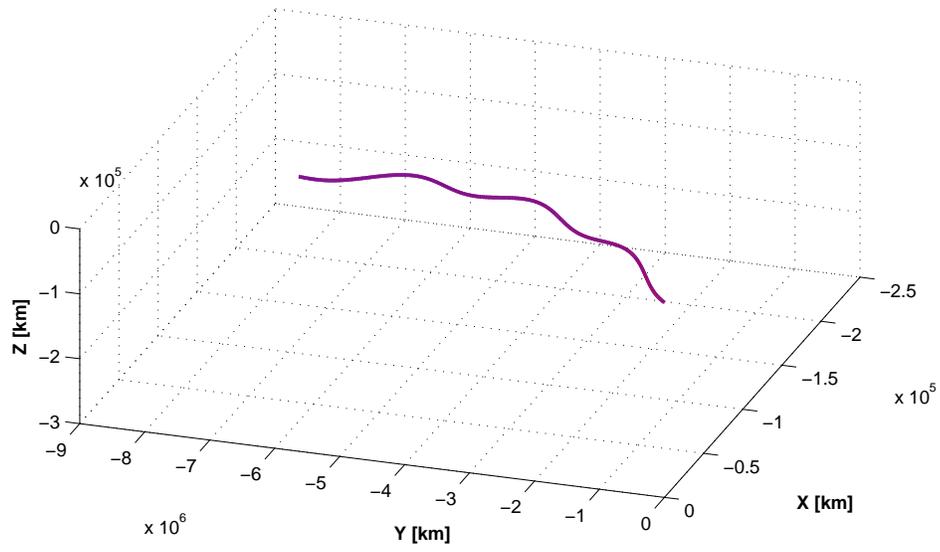}
\end{center}
\caption{Space trajectory of the Earth with respect to the pulsars during three days. At this scale the ideal and the reconstructed curves are indistinguishable. }
\label{fig:eartha}
\end{figure}

The results are shown in Figure \ref{fig:eartha} where the reconstructed spatial trajectory is compared with the one
determined by the ICRS Ephemerides of the chosen observatory. The scale of the figure does not permit to appreciate the
differences between the two trajectories. Actually this application of the method is purely indicative. TEMPO2 has of course not been designed for our purposes, so the sampling of the data each 10 seconds may introduce some additional uncertainty; moreover, as we stressed in \cite{pulsararc10} referring to the Geometric Dilution Of Precision (GDOP), a crucial role in minimizing the uncertainty is played by the geometry of the sources: the uncertainty is minimized  when the volume spanned by the sources directions is maximized.

We stress once more the demonstrative purpose of our work, which has led us to disregard a series of aspects that should be taken into account for a real positioning system. We have for instance assumed that all four pulsars of our example are simultaneously visible in the sky of Parkes at any moment, which is not the case. We could have chosen a set of circumpolar pulsars in order to achieve the continuous visibility condition. In that case however the sources being located in the same region of the sky would have spoilt the quality of the positioning for pure geometric reasons. In a real system we would need redundancy, considering more than four sources (as it is the case for the terrestrial GPS) so that at least four are always above the horizon. The system would use several quartets and would have to smoothly pass from one to an another when some of the stars sets down or comes up. In any case the simultaneous use of more than one set of four sources would also allow to obtain the positioning by way of an averaging process, even in the case of a practically pointlike receiver, such as a spacecraft.
Another problem would arise from the fact that the definition of the sequence of the arrival times, getting rid of perturbations and noise, requires an integration time that could be too long for the approximation of a linear worldline to hold. However if we consider the case of the motion of the Earth, we see that the deviation of the worldline from the rectilinear trend in 10 seconds is in the order of 1 part in 10$^{10}$, which means that there is reasonable room for the integration. With all that we thought that the 1 microsecond $\sigma$ noise we introduced can reasonably well account for most short time disturbances in the arrival times.


\section{Conclusions}\label{sec:conc}

We studied  the possibility of using  sources which emit pulsating electromagnetic signals for positioning purposes and, in particular, we focused on a fully relativistic approach that we introduced in a previous work,
which allows positioning with respect to an arbitrary event in flat spacetime. This approach is based on the definition of a null frame,  by means of the four-vectors associated to the signals in the inertial reference frame where the sources are at rest, which, in turn, are determined by the emission directions and the frequencies of the pulsating signals.

The procedure for positioning determination rests upon the hypothesis that the receiver's world-line is a straight line during a proper time interval corresponding to the reception of a limited number of pulses, which holds true if the effects of its acceleration are negligibly small during that time. This is indeed true for any solid system when the time span is only a fraction of a second, of the order of, say, one hundredth or less. Of course also the space-time curvature has to be conveniently small in order for the hypothesis to hold, but again this is the case in the Solar system and for the conditions of the simulation.

We tested the reliability of our procedure by means of numerical simulations, on the basis of the definition of null frames where the sources can be thought of as pulsars.

As a simple example, we showed that our algorithms can be used to reproduce the spacetime trajectory of an user at rest, with an accuracy that is of the same order as the accuracy with which the data are known.
After this initial calibration, we considered a more interesting case, that is the motion of the Earth. In particular, on using a simulating plug-in of the TEMPO2 software, we sampled the times of arrival of the signals from a given set of pulsars, expected from an observation session at a specific location on the Earth, which in our case is the Parkes observatory in Australia. By collecting data which simulate an observation of about three days, we determined the trajectory of the observatory due to the combined rotation and revolution motion of the Earth.  Then, we compared the reconstructed world-line with the one obtained by the ephemerides. The comparison was only for qualitative purposes since the use of TEMPO2 intermittently and for such a long (from the view point of our method) time may introduce some additional uncertainty; furthermore the choice and the use we made of pulsars corresponds to an idealized situation. Our method would in general not be used for precision astrometry, but rather for navigation and most likely the emitters would be artificial sources. Artificial pulsating sources would provide much higher frequencies than pulsars and much higher signal intensities. This would mean smaller antennas and receivers, then more portable devices. On the other side artificial "pulsars" could not be in anyway "fixed" into the sky and we would have to know very well their worldlines. Direction cosine will be functions of time as well as the frequencies received by an observer at rest. With all that the method and the algorithm would remain essentially the same.

In any case our preliminary results show the feasibility of the use of pulsating sources for positioning purposes, in a fully relativistic framework. Of course, in order to deal with true pulsars as well as artificial signals further steps are necessary to study and define many technological aspects, if we wish to  include an increasing number of practical and technical details of the acquisition of the signals and the subsequent data processing. It is important to repeat and stress that our approach can be applied to artificial sources, such as pulsating sources on board of spacecrafts or celestial bodies in the Solar System: to this end, our procedure should be generalized  to take into account the fact that the sources are not fixed, but follow closed space orbits, and that signals propagate in a gravitational field. In any case, both the usage of pulsars signals and that of artificial sources require further investigations, which seem worthwhile since, as we have showed here, in any case the self-sufficiency of the method proves to be its main advantage, while giving quite acceptable results.


\section*{Acknowledgements}

\label{sec:ack} 

We are grateful to Dr. Andrea Possenti for helpful discussions about pulsar timing and the use of the TEMPO2 software.

Our research has been supported by Piemonte local government within the
MAESS-2006 project "Development of a standardized modular platform for
low-cost nano- and micro-satellites and applications to low-cost space
missions and to Galileo" and by ASI.

\protect \pagebreak


{\normalsize}

{\normalsize}

{\normalsize}


\protect \pagebreak

\end{document}